# Electronic states of PrCoO$_3$: X-ray photoemission spectroscopy and LDA+U density of states studies

S. K. Pandey, Ashwani Kumar[a], S. M. Chaudhari, and A. V. Pimpale

UGC-DAE Consortium for Scientific Research, University Campus, Khandwa Road, Indore 452 017, India

[a]School of Physics, Devi Ahilya University, Khandwa Road, Indore 452 017, India

E-mail: sk_iuc@rediffmail.com

**Abstract**

Electronic states of PrCoO$_3$ are studied using X-ray photoemission spectroscopy. Pr 3d$_{5/2}$ core level and valence band (VB) were recorded using Mg K$_\alpha$ source. The core level spectrum shows that the 3d$_{5/2}$ level is split into two components of multiplicity 4 and 2, respectively due to coupling of the spin states of the hole in 3d$_{5/2}$ with Pr 4f holes spin state. The observed splitting is 4.5 eV. The VB spectrum is interpreted using density of states (DOS) calculations under LDA and LDA+U. It is noted that LDA is not sufficient to explain the observed VB spectrum. Inclusion of on-site Coulomb correlation for Co 3d electrons in LDA+U calculations gives DOS which is useful in qualitative explanation of the ground state. However, it is necessary to include interactions between Pr 4f electrons to get better agreement with experimental VB spectrum. It is seen that the VB consists of Pr 4f, Co 3d and O 2p states. Pr 4f, Co 3d and O 2p bands are highly mixed indicating strong hybridization of these three states. The band near the Fermi level has about equal contributions from Pr 4f and O 2p states with somewhat smaller contribution from Co 3d states. Thus in the Zaanen, Sawatzky, and Allen scheme PrCoO$_3$ can be considered as charge transfer insulator. The charge transfer energy $\Delta$ can be obtained using LDA DOS calculations and the Coulomb-exchange energy U′ from LDA+U. The explicit values for PrCoO$_3$ are $\Delta$ = 3.9 eV and U′ = 5.5 eV; the crystal field splitting and 3d bandwidth of Co ions are also found to be 2.8 and 1.8 eV, respectively.

**Introduction**

Perovskite type transition metal oxides have been of much interest for more than fifty years. Recently, research activities have been intensified in this class of materials due to the emergence of exotic properties like charge-disproportionation, charge ordering (CO), orbital ordering (OO), phase separation, colossal magneto resistance (CMR) etc. [1]. The interplay between the on-site and inter-site Coulomb interaction, the charge transfer energy, the hybridization strength between the cation $3d$ and oxygen $2p$ states and the crystal field splitting for the $d^n p^m$ configuration of the $MO_6$ (M = $3d$ transition metal ion) octahedron control the ground state electronic structure and magnetic and transport properties of these perovskites.

Cobaltates $ACoO_3$, where A is the rare earth element, forms an interesting class of compounds in the perovskite family. These compounds show insulator to metal transition with increase in temperature [2]. It is believed that such transition occurs due to thermally driven spin state transition of $Co^{3+}$ ions [3]. Goodenough originally proposed that, instead of an intermediate spin (IS) $t_{2g}^5 e_g^1$ state that could undergo a disproportionation reaction $2t_{2g}^5 e_g^1 \leftrightarrow t_{2g}^5 e_g^0 + t_{2g}^5 e_g^2$, the system would find it energetically more favourable to stabilize high-spin (HS) $t_{2g}^4 e_g^2$ states where the nearest neighbours can be in the low-spin (LS) $t_{2g}^6 e_g^0$ state; such an ordered configuration could achieve the extra covalent bonding in a LS $CoO_6$ site without costing the energy required to create $Co^{4+}$ and $Co^{2+}$ ions. In 1995, Potze et al. revived the interest in these compounds by giving three-spin-state (LS-IS-HS) model for $LaCoO_3$. In this mechanism $Co^{3+}$ ions are thermally activated to the IS state from LS state at 100 K, and then from IS state to a mixed state of IS and HS around 500 K.

Electronic structures of these compounds depend on the effective d-d Coulomb and exchange energy (U′) required for $d_i^n d_j^n \leftrightarrow d_i^{n-1} d_j^{n+1}$ (where $i$ and $j$ are the labels for transition metal site) and the charge transfer energy (Δ) required for $d_i^n \rightarrow d_i^{n+1} \underline{L}$, where $\underline{L}$ denotes a hole in the anion valence band. Zaanen et al. [4] proposed a model for describing band gaps ($E_g$) and electronic structures of transition metal (TM) compounds. They divided these compounds into four categories, two of them are relevant in the context of cobaltates (1) Mott-Hubbard (MH) and (2) charge-transfer (CT) type, depending on the values of U′ and Δ. In the case of MH type U′ < Δ and $E_g$ is proportional to U′, whereas for CT type compounds U′ > Δ and $E_g$ is proportional to Δ. In addition to these two parameters electronic structures of such compounds depend on hybridization of the TM 3d and ligand 2p states. Generally Δ is equal to the difference between centre of gravity (CG) of d and p bands when the correlation between d electrons (i.e. U′) is not considered [1]. In this case systems show metallic behavior because of finite d symmetric density of states at the Fermi level and p bands lie below d bands. On invoking U′, a band gap is created at the Fermi level and the system shows insulating behaviour. In the case of MH type compounds, electrons near the Fermi level are of d character, whereas for the CT type compounds they are of p character.

It is seen that in many TM compounds LMM Auger spectrum shows sharp atomic like features. Sawatzky [5] showed that d-d correlation can give rise to such effects and the sharp line appears on lower kinetic energy (KE) side of the broad band arising from the self-convolution of the 3d band involved. The energy separation between the peaks of the sharp line and the broad band is a measure of the Coulomb correlation energy U′. Experimentally such interpretation has been used for $LaFeO_3$, $LaCoO_3$ and $LaNiO_3$ [6]; however, such an interpretation has also been questioned in $LaCoO_3$ [7].

We report here a study of electronic states of $PrCoO_3$ by room temperature X-ray photoemission spectroscopy (XPS) and electronic structure calculations using LDA+U method [8]. Although considerable work is seen in literature on electronic states of $LaCoO_3$, hardly any work exists on $PrCoO_3$. In contrast to rhombohedral $LaCoO_3$, $PrCoO_3$ is orthorhombic. Its ground state is paramagnetic insulator with low spin state [9]. Two main differences can be readily seen in the $LaCoO_3$ and $PrCoO_3$ compounds: (a) 4f states of $La^{3+}$ ion are unoccupied and that of $Pr^{3+}$ contain two electrons, (b) ionic size of $Pr^{3+}$ is smaller than the ionic size of $La^{3+}$. Differences in the ionic size result in changed lattice distortion and directly affect the hybridization and thus the bandwidth. This also affects the crystal-field splitting energy and spin state transition, since the latter occurs due to competition between crystal-field energy and Hund's coupling energy. Due to presence of two electrons in 4f states of $Pr^{3+}$ ion one expects the contribution of these states in the valence band. As mentioned above, the inclusion of on-site Coulomb and exchange interactions completely change the energy positions of d and p bands; therefore it would be interesting to see their effects especially on the occupied 4f band. We also look at the possibility of finding the values of parameters $\Delta$ and $U'$ using LDA and LDA+U density of states (DOS), respectively. This will help in understanding whether the compound under study is MH or CT type as per the scheme of Zaanen, Sawatzky, and Allen (ZSA) [4].

The experimental core level spectrum shows that the Pr $3d_{5/2}$ level is split into two components of multiplicity 4 and 2, respectively due to coupling of the spin states of the hole in $3d_{5/2}$ with that of Pr 4f holes. The observed splitting is 4.5 eV. The valence band (VB) spectrum of $PrCoO_3$ consists of O 2p, Co 3d and Pr 4f states as revealed by calculated occupied (DOS). Within LDA formulation, DOS of occupied states near the Fermi level is a mixture of Pr 4f and Co 3d states. On invoking LDA+U, energy position as well as DOS of all the bands are affected in such a way that the occupied O 2p band is shifted near the Fermi level and Co 3d and Pr 4f bands move away from it, so that the band near the Fermi level is mainly of O 2p and Pr 4f character than Co 3d character. The calculated values of $U'$ and $\Delta$ are 5.5 eV and 3.9 eV, respectively, indicating that $PrCoO_3$ is of CT type compound. The crystal field splitting and 3d bandwidth of Co ions are found to be 2.8 and 1.8 eV, respectively.

**Experimental**

Single-phase polycrystalline powder samples of $PrCoO_3$ and $Pr_{0.8}Ca_{0.2}CoO_3$ were prepared by combustion method [10]. They were characterized by powder X-ray diffraction (XRD) and resistivity techniques. Lattice parameters obtained from Rietveld analysis match well with literature. Resistivity measurements revealed that both the samples are insulating. The XPS measurements were carried out using photoelectron spectrometer equipped with an OMICRON electron analyzer (model EA125). All the Core levels, Auger and valence band spectra reported in the present work were carried out using Mg $K_\alpha$ radiation at 50 eV pass energy of the spectrometer. The measured resolution was about 0.85 eV. The sample was mounted in the form of a compressed hard pellet and it was scraped uniformly by diamond file before carrying out the measurements. Final spectrum was taken only when the feature coming from carbon contamination of the surface merged in the spectrum background. The Fermi level ($E_f$) was aligned by recording the VB spectrum of in situ cleaned gold foil. In order to check for any deviation in $E_f$, VB spectra of gold were recorded immediately after taking the photoemission spectrum of each compound under present study. Vacuum of the chamber during the experiment was $5\times10^{-10}$ Torr.

**Computational details**

Spin unpolarized electronic structure calculations were carried out using LMTART 6.52 [11]. For calculating charge density, full-potential LMTO method working in plane wave representation was used. In the calculation we have used the muffin-tin radii of 3.521, 2.03, and 1.728 a.u. for Pr, Co, and O sites, respectively corresponding to lattice parameters of $PrCoO_3$. The charge density and effective potential are expanded in spherical harmonics up to $l = 6$ inside the sphere and in a Fourier series in the interstitial region. The initial basis set included 6s, 5p, 5d, and 4f valence and 5s semicore orbitals of Pr; 4s, 4p, and 3d valence and 3p semicore orbitals of Co and 2s and 2p valence orbitals of O. The exchange correlation functional of the density functional theory was taken after Vosko, Wilk, and Nussair and the generalized-gradient approximation scheme of Perdew et al. [12] was also invoked. (6, 6, 6) divisions of the Brillouin zone along three directions for the tetrahedron integration were used to calculate the DOS. Self-consistency was achieved by demanding the convergence of the total energy to be smaller than $10^{-4}$ Ry/cell.

LDA+U gives better explanation for electronic properties of the systems containing transition metal or rare-earth metal ions or both with partially filled d (and/or f) orbitals whereas LDA is useful for the systems consisting of delocalized s, p electrons [8]. We used both LDA and LDA+U methods to calculate DOS of $PrCoO_3$. In the LDA interaction potential depends on the average local electronic density and hence the occupied and unoccupied orbitals are degenerate. Due to this LDA fails to create gap at the Fermi level and the calculated DOS shows metallic behaviour. This discrepancy is resolved in the LDA+U method where interaction potential depends on the orbital degrees of freedom and hence the eigenvalues of occupied and unoccupied orbitals are different. Thus the method is capable of creating gap at the Fermi level and giving the correct ground state of the system. In the LDA+U method three input parameters are required for 3d electrons. They are Slater integrals $F^0$, $F^2$, and $F^4$. These integrals are directly related with on-site Coulomb interaction (U) and exchange interaction (J) by relations $U = F^0$, $J = (F^2 + F^4) / 14$, and $F^4/ F^2 \sim 0.625$ [13]. We have tried several values of U (1.5 eV to 8.5 eV in steps of 1.0 eV) and J (0.3 to 1.2 eV in steps of 0.1 eV). The calculated DOS with U = 3.5 eV and J = 1.0 eV were the best representative of the experimental VB spectrum. Moreover, these values of U and J gave the ground state with lowest energy. This value of J is close to the earlier reported value for CoO [14]. Four input parameters are also required when the on-site Coulomb and exchange interactions between 4f electrons are considered in the calculations. They are Slater integrals $F^0$, $F^2$, $F^4$, and $F^6$ (the same symbol F is used for simplicity as in the 3d case; this should cause no confusion). These integrals are directly related with on-site Coulomb interaction (U) and exchange interaction (J) by relations $U = F^0$, $J = (286F^2 + 195F^4 + 250F^6) / 6435$ [8]. For rare-earth elements $F^4/F^2 = 0.624$ and $F^6/F^2 = 0.448$ [15]. Judd et al.[15] have also given the value of $F^2 = 8.647$ eV for $Pr^{3+}$ ion. We have taken this value of $F^2$ in our calculations and varied $F^0$ from 1.5 to 9.5 eV in steps of 0.5 eV. The values of U and J for Co 3d and Pr 4f electrons obtained from the above relations that give the DOS which is the best representative of the experimental VB spectrum are given in table 1. The DOS calculated using these values is used for discussion in the next section. It is interesting to note that the value of U = 3.5 eV, giving DOS which is the best representative of the experimental VB spectrum, is approximately equal to the separation of two Co $L_3$ $M_{4,5}$ $M_{4,5}$ Auger peaks (Fig. 1). It will be interesting to see whether this observation is unique for $PrCoO_3$ or it is a common feature for all such compounds.

**Results and discussions**

Figure 2 shows satellite subtracted and fitted core level spectrum of Pr $3d_{5/2}$ states. It contains two peaks, one at 928.5 eV and another at 933.0 eV. The BEs of Pr $3d_{5/2}$ and Co 2s states for neutral Pr and Co atoms are 928.8 eV and 925.1 eV, respectively. These values are expected to increase for $Pr^{3+}$ and $Co^{3+}$ ions. The theoretical ratio of cross sections of Pr $3d_{5/2}$ to Co 2s is about 7.8 for Mg $K_\alpha$ radiation [16]. The integrated intensity under the experimental peak at 933.0 eV is 1.93 times that under the peak at 928.5 eV. If one of the peaks is considered as arising from Co 2s and another from Pr $3d_{5/2}$, the observed intensities are in contradiction with the known cross sections for these peaks. We thus interpret both the peaks as arising from Pr $3d_{5/2}$. There might be some Co 2s contribution in the experimental data, however it is not resolvable. The two peaks can be understood by considering exchange splitting of Pr $3d_{5/2}$ level. In the photoelectron emission process one hole is created in the Pr $3d_{5/2}$ state. Spin state of this hole couples with the spin state of Pr 4f holes and results in exchange splitting of Pr $3d_{5/2}$ state. The multiplicity of the doublet thus created is 4 and 2 and hence the ratio of the integrated intensity of the two peaks should be 2:1, which is close to the observed value. The energy separation of the doublet formed by exchange splitting is the energy difference between these two peaks, which is 4.5 eV. We have also observed similar two peaks in $PrFeO_3$ compound [10].

In figure 3 the background subtracted VB spectrum of $PrCoO_3$ is shown by open circles. At the Fermi level, the curve exhibits a very low intensity i.e. a shape corresponding to the tail of decreasing photoemission signal. Such a small intensity gives an impression that the sample is metallic at room temperature. But our resistivity data show insulating behaviour. It is well established that a small intensity at Fermi level may be seen even in the insulating samples due to finite resolution of the spectrometer. Such situation has been noted by other workers as well [6, 17]. Two peaks, labeled as 1 and 2, are clearly visible in the VB spectrum, sharper one at 1.6 eV and broader one at 5.2 eV. To identify the contribution of different bands to these peaks we have calculated partial DOS. Figure 4 shows the DOS of $PrCoO_3$ including the partial DOS of Pr 4f, Co 3d, and O 2p calculated under local density approximation (LDA). It is evident from the figure that the occupied DOS can be divided into three distinct bands labeled by 1′, 1 and 2. Band 1′ mainly consists of Co 3d and Pr 4f states. Band 1 has contribution mainly from O 2p states and band 2 has mixed O 2p and Co 3d character. These three bands are well separated and the separation between them is greater than the experimental resolution. Thus the experimental VB spectrum with only two peaks cannot be accounted for by using these DOS. Moreover, there is finite DOS at the Fermi level due to contribution from Pr 4f and Co 3d states indicating that the $PrCoO_3$ should be conducting, which is contrary to the experimentally observed insulating behaviour. Such failure of LDA in explaining transport properties of transition metal oxides from Fe onwards has been seen in literature. Hence we used LDA+U method to calculate the DOS.

Figure 5 shows the calculated DOS of PrCoO$_3$ using LDA+U method ignoring Coulomb and exchange interactions between Pr 4f electrons but including them for Co 3d electrons. Upper part of the figure shows total DOS of the system. It shows that all bands are affected in such a way that bands of metal ions (i.e. Co$^{3+}$ and Pr$^{3+}$) shift to higher BE and that of ligand ion (i.e. O$^{2-}$) to lower BE. Total occupied DOS consists of Pr 4f, Co 3d, and O 2p bands. It can be divided into three bands indicated by 1′, 1 and 2 in the figure. Band 1′ lies just near the Fermi level, band 1 between 1.5 eV to 3.5 eV and band 2 between 3.6 eV to 6.6 eV. The bandwidths of band 1 and 2 are about 2.0 eV and 3.0 eV, respectively, whereas band 1′ is very sharp. The separation between band 1′ and 1 is about 1.1 eV, which is more than the experimental resolution. From the figure it is evident that bands 1 and 2 have highly mixed character of Co 3d and O 2p electronic states indicating strong hybridization between them. This fact is also indicated by the total occupancy of the 3d orbitals. Calculations give this to be 6.8, which is 0.8 larger than the nominal occupancy 6 of the 3d orbitals in Co$^{3+}$. For LaCoO$_3$ total occupancy of 3d orbitals found by Korotin et al. [3] is 7.2, a value greater than our value of 6.8 in PrCoO$_3$. This may be due to the fact that PrCoO$_3$ is more distorted than LaCoO$_3$, thereby decreasing the overlap of O 2p and Co 3d orbitals. If we look at the top of the VB (excluding the contribution from Pr 4f states) we find that it is a mixture of dominant O 2p orbitals with Co $t_{2g}$ orbitals. Such behaviour of top of the VB would indicate that PrCoO$_3$ is a charge transfer type compound under the ZSA scheme [4]. The top of the VB of CuO, which is a well-known CT type compound, also shows similar nature [8].

A simple way to understand the behaviour of electronic states in PrCoO$_3$ and similar compounds is shown schematically in figure 6. This figure is a modified form of the one given by Imada et al. [1]. The left side of this figure gives calculated DOS under LDA and the right side shows the expected change when LDA+U is considered. When on-site Coulomb and exchange interaction (U′) is not considered i.e. under LDA, the top of the valence band has only d character and p band lies just below d band. The separation between their CG is the charge transfer energy Δ. On including U′, p band shifts up closer to the Fermi level and d band shifts down away from the Fermi level in such a way that occupied d band lies below p band. The energy difference of CG between the occupied and unoccupied d bands is then equal to U′. As can be seen from the comparison of figures 4 and 5, on inclusion of U′ band 1 of Co 3d states shifts to higher BE by 1.5 eV while the CG of band 2 remains almost unchanged. However, DOS of the band 2 grows at the cost of DOS of band 1. Both bands of O 2p states show opposite behaviour to Co 3d bands. Band 1 of O 2p states shifts to lower BE by 0.4 eV with increased DOS, whereas position band 2 remains almost unchanged but its DOS decreases. On considering on-site Coulomb and exchange interaction of 3d electrons readjustment of the BE positions and DOS of O 2p and Co 3d bands occurs in such a way that top of the VB has dominating 2p character which is the characteristic of CT type system. It is thus seen that LDA and LDA+U can be used to extract the values of parameters Δ and U′, respectively. For $PrCoO_3$, Δ is found to be about 3.9 eV. This value is close to the value obtained by Chainani et al. [6] for $LaCoO_3$. The value of U′ for $PrCoO_3$ is about 5.5 eV. This value is closer to the value obtained by Saitoh et al. [17] by analyzing Co 2p core level XPS spectrum of $LaCoO_3$. The values of U′ and Δ thus obtained are compatible for CT type insulator as U′ > Δ. Hence LDA and LDA+U method can together be used to categorize the materials based on ZSA scheme. It is interesting to note that the value of U′ is not simple sum of Coulomb and exchange contributions; instead in the present compound it is $U + 2J$. This has also been observed in other systems, e.g. in Mn alloy it is $U + 4J$ [18].

Thus we see that inclusion of on-site Coulomb and exchange interactions between Co 3d electrons only, gives a considerable understanding of the system. However, the experimental VB spectrum with only two peaks cannot be accounted for by using these DOS. Moreover, there is significant DOS at the Fermi level due to contribution from Pr 4f. As $PrCoO_3$ is insulating, there should be a gap at the Fermi level. Thus, only on-site Coulomb and exchange interaction between Co 3d electrons is not sufficient to give the correct ground state for $PrCoO_3$. This deviation of the calculated DOS from the observed behaviour is mainly due to non-inclusion of the on-site Coulomb and exchange interaction between Pr 4f electrons. Pr 4f electrons are more localized than Co 3d electrons and the on-site Coulomb and exchange interaction is expected to be more than that for Co 3d electrons. Thus it may be anticipated that Pr 4f band (i.e. band 1′) will merge with band 1 on including on-site interaction, thereby creating a gap at the Fermi level.

To check this explicitly, we calculated the DOS including on-site Coulomb and exchange interactions for both Co 3d and Pr 4f electrons. Figure 7 shows the calculated DOS including the partial DOS of $PrCoO_3$. Total occupied DOS is shown in the upper part of the figure. It consists of Pr 4f, Co 3d, and O 2p states. Closer inspection of the total DOS near the Fermi level shows a clear jump just above the Fermi level. DOS just below the Fermi level is about 25 times lower than that just above it. The occupied DOS can be divided into two bands indicated by 1 and 2 in figure 7. Band 1 lies between 1.2 eV to 3.4 eV and band 2 between 3.7 eV to 6.5 eV. As can be seen from the comparison of figures 5 and 7, on inclusion of interaction between 4f electrons, band 1 shifts near to the Fermi level by 0.3 eV making it broader, whereas the position and width of band 2 remains almost unchanged. The peak position and bandwidth of band 1 is 1.6 eV and 2.2 eV, respectively. This peak position is almost the same as the position of peak 1 of VB spectrum (Fig. 3). Thus the calculated DOS clearly represents the experimental VB spectrum.

The calculated partial DOS indicates that Pr 4f states contributes only in peak 1 of VB spectrum whereas Co 3d and O 2p states contribute to both the peaks. Region near the Fermi level has largest contribution from Pr 4f states and least from the Co 3d states. Peak 1 has highly mixed character of Pr 4f, O 2p and Co 3d states and peak 2 has highly mixed character of only O 2p and Co 3d states. The mixing of these states is an indication of their strong hybridization. This fact is indicated by the total occupancies of 3d and 4f orbitals. Calculations give the total occupancy of 4f orbitals to be 2.4, which is 0.4 larger than the nominal occupancy of these orbitals in $Pr^{3+}$. The total occupancy of 3d orbitals does not change on inclusion of interaction between 4f electrons. Such values of total occupancies of the Pr 4f and Co 3d orbitals may indicate the transfer of charge from O sites to Pr and Co sites. A simple way to check the contribution of Pr 4f in peak 1 experimentally is to change the relative concentration of Pr by doping. We replaced 20% Pr by Ca and recorded the VB spectrum. Closed circles in figure 3 indicate the VB spectrum of $Pr_{0.8}Ca_{0.2}CoO_3$. It also consists of two peaks, 1 and 2 around 1.6 and 5.2 eV, respectively. The intensity of the peak 1 decreases with respect to peak 2 indicating the contribution of Pr 4f states in peak 1.

On inclusion of on-site Coulomb and exchange interaction between Pr 4f electrons the energy difference of the CG of occupied and unoccupied 3d bands remains almost unchanged giving the same value of U′ as obtained earlier indicating that $PrCoO_3$ is indeed a CT insulator. Besides U′ and Δ, the crystal field-splitting energy and bandwidth W of transition metal 3d band are also important physical parameters for cobaltates. Crystal field-splitting energy governs the temperature dependent spin state transitions. It is the energy difference between the CG of $t_{2g}$ and $e_g$ orbitals, which is 2.8 eV for $PrCoO_3$. This value, though larger than that of $LaCoO_3$ [19], is compatible with higher spin state transition temperature of $PrCoO_3$ in comparison to $LaCoO_3$. W is the bandwidth of 3d band of transition metal when on-site Coulomb and exchange interaction is absent. It is the measure of delocalization of electrons. The competition between W and U′ decides the physical properties of the system. For $PrCoO_3$, W ~ 1.8 eV was obtained from LDA DOS calculations.

**Conclusions**

Electronic states of $PrCoO_3$ have been studied using XPS core level, valence band (VB); LDA and LDA+U DOS calculations. Pr $3d_{5/2}$ core level showed the spin coupled exchange splitting of 4.5 eV. It is found to be necessary to include on-site Coulomb and exchange interactions between Pr 4f electrons in addition to those between Co 3d electrons to get better agreement with the experimental VB spectrum. VB consisted of highly mixed Pr 4f, Co 3d and O 2p states. Band near the Fermi level have dominating O 2p and Pr 4f character than Co 3d character. The values of charge transfer energy Δ = 3.9 eV and Coulomb and exchange energy U′ = 5.5 eV were found, indicating that $PrCoO_3$ is CT insulator in conformity with Zaanen, Sawatzky and Allen scheme. The crystal field splitting energy was found to be 2.8 eV and the bandwidth of Co 3d band was obtained to be 1.8 eV.

**Acknowledgements**

We would like to acknowledge R. Rawat; N. P. Lalla and S. Bhardwaj for help in resistivity and XRD measurements respectively. SKP thanks UGC-DAE CSR for financial support. AK thanks CSIR for senior research associate position (pool scheme) of Government of India.

**References**

[1] M. Imada, A. Fujimori, and Y. Tokura, Rev. Mod. Phys. **70**, 1039 (1998)

[2] S. Yamaguchi, Y. Okimoto, and Y. Tokura, Phys. Rev. B **54**, R11022 (1996)

[3] J. B. Goodenough, J. Phys. Chem. Solids **6**, 287 (1957); R. H. Potze, G. A. Sawatzky, and M. Abbate, Phys. Rev. B **51**, 11501 (1995); M. A. Korotin, S. Yu. Ezhov, I. V. Solovyev, V. I. Anisimov, D. I. Khomskii, and G. A. Sawatzky, Phys. Rev. B **54**, 5309 (1996)

[4] J. Zaanen, G. A. Sawatzky, and J. W. Allen, Phys. Rev. Lett. **55**, 418 (1985)

[5] G. A. Sawatzky, Phys. Rev. Lett. **39**, 504 (1977)

[6] A. Chainani, M. Mathew, and D. D. Sarma, Phys. Rev. B **46**, 9976 (1992); A. Chainani, M. Mathew, and D. D. Sarma, Phys. Rev. B **48**, 14818 (1993); S. R. Barman, A. Chainani, and D. D. Sarma, Phys. Rev. B **49**, 8475 (1994)

[7] D. D. Sarma and P. Mahadevan, Phys. Rev. Lett. **81**, 1658 (1998)

[8] V. I. Anisimov, F. Aryasetiawan, and A. I. Lichtenstein, J. Phys.: Condens. Matter **9**, 767 (1997)

[9] S. Tsubouchi, T. Kyomen, M Itoh, and M Oguni, Phys. Rev. B **69**, 144406 (2004)

[10] S. K. Pandey, R. Bindu, P. Bhatt, S. M. Chaudhari, and A. V. Pimpale, Physica B, **365**, 45 (2005)

[11] S. Y. Savrasov, Phys. Rev. B **54**, 16470 (1996)

[12] J. P. Perdew, K. Burke, and M. Ernzerhof, Phys. Rev. Lett. **77**, 3865 (1996)

[13] V. I. Anisimov, I. V. Solovyev, M. A. Korotin, M. T. Czyzyk, and G. A. Sawatzky, Phys. Rev. B **48**, 16929 (1993)

[14] V. I. Anisimov, J. Zaanen, and O. K. Andersen, Phys. Rev. B **44**, 943 (1991)

[15] B. R. Judd and I. Lindgren, Phys. Rev. **122**, 1802 (1961)

[16] J. J. Yeh and I. Lindau, At. Data Nucl. Data Tables **32**, 1 (1985)

[17] T. Saitoh, T. Mizokawa, A. Fujimori, M. Abbate, Y. Takeda, M. Takano, Phys. Rev. B **55**, 4257 (1997)

[18] D. van der Marel, G. A. Sawatzky, and F. U. Hillebrecht, Phys. Rev. Lett. **53**, 206 (1984)

[19] I. A. Nekrasov, S. V. Streltsov, M. A. Korotin, and V. I. Anisimov, Phys. Rev. B **68**, 235113 (2003)

**Figure captions**

**Figure 1.** Co $L_3 M_{4,5} M_{4,5}$ Auger spectrum of $PrCoO_3$.

**Figure 2.** Pr $3d_{5/2}$ X-ray photoemission spectrum. Two peaks are arising from spin coupled exchange interaction with 4f holes. Open circles represent the experimental spectrum and solid lines correspond to the fitted curves including the background.

**Figure 3.** Background subtracted valence band spectra of $PrCoO_3$ (open circles) and $Pr_{0.8}Ca_{0.2}CoO_3$ (closed circles).

**Figure 4.** DOS of $PrCoO_3$ calculated using LDA method (different Y-scales are used for different partial DOS for the sake of clarity).

**Figure 5.** DOS of $PrCoO_3$ calculated using LDA+U method with on-site Coulomb and exchange interaction between Co 3d electrons included (different Y-scales are used for different partial DOS for the sake of clarity).

**Figure 6.** Schematic diagram of DOS for CT insulator; left corresponding to LDA and right corresponding to LDA+U.

**Figure 7.** DOS of $PrCoO_3$ calculated using LDA+U method with on-site Coulomb and exchange interaction between Co 3d and Pr 4f electrons included (different Y-scales are used for different partial DOS for the sake of clarity).

Table 1: The best values of on-site Coulomb interaction, U and exchange interaction, J for Co 3d and Pr 4f electrons corresponding to the experimental valence band spectrum of PrCoO$_3$.

| method of calculation | on-site Coulomb-exchange interaction for | U (eV) | J (eV) |
|---|---|---|---|
| LDA | - | 0.0 | 0.0 |
| LDA+U | Co 3d electrons | 3.5 | 1 |
|  | Pr 4f electrons | 3.5 | 0.698 |

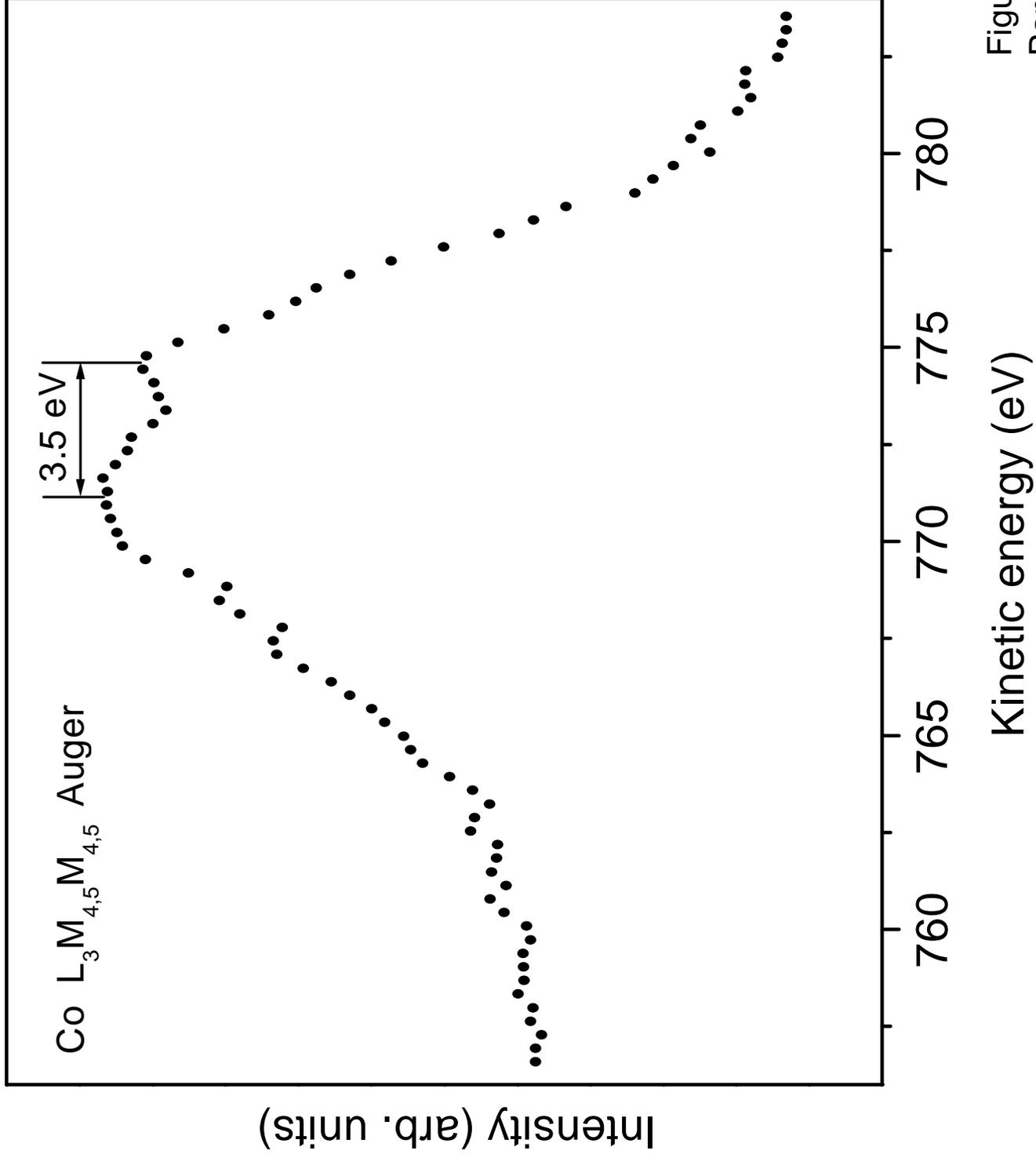

Figure 1
Pandey et al.

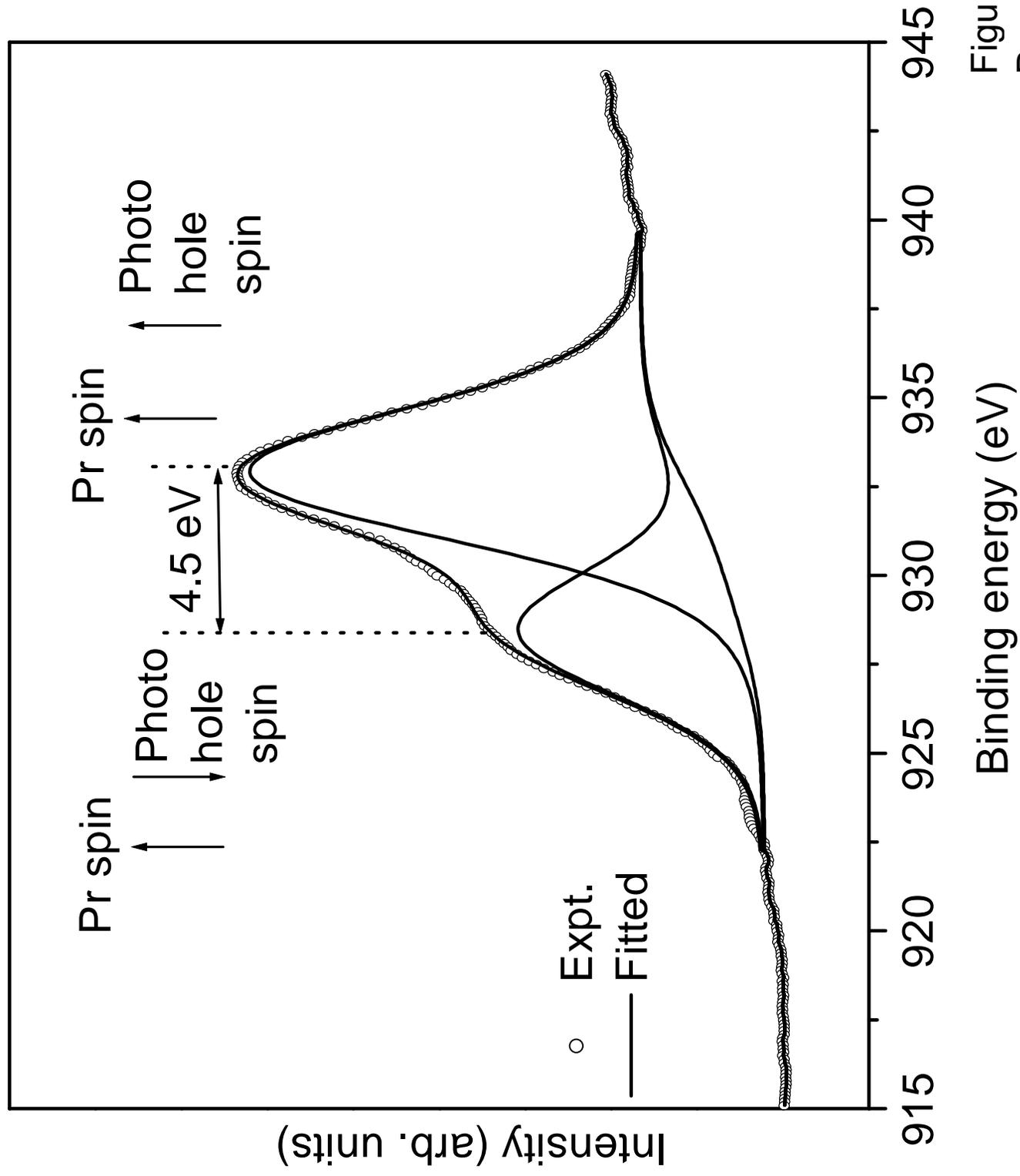

Figure 2
Pandey et al.

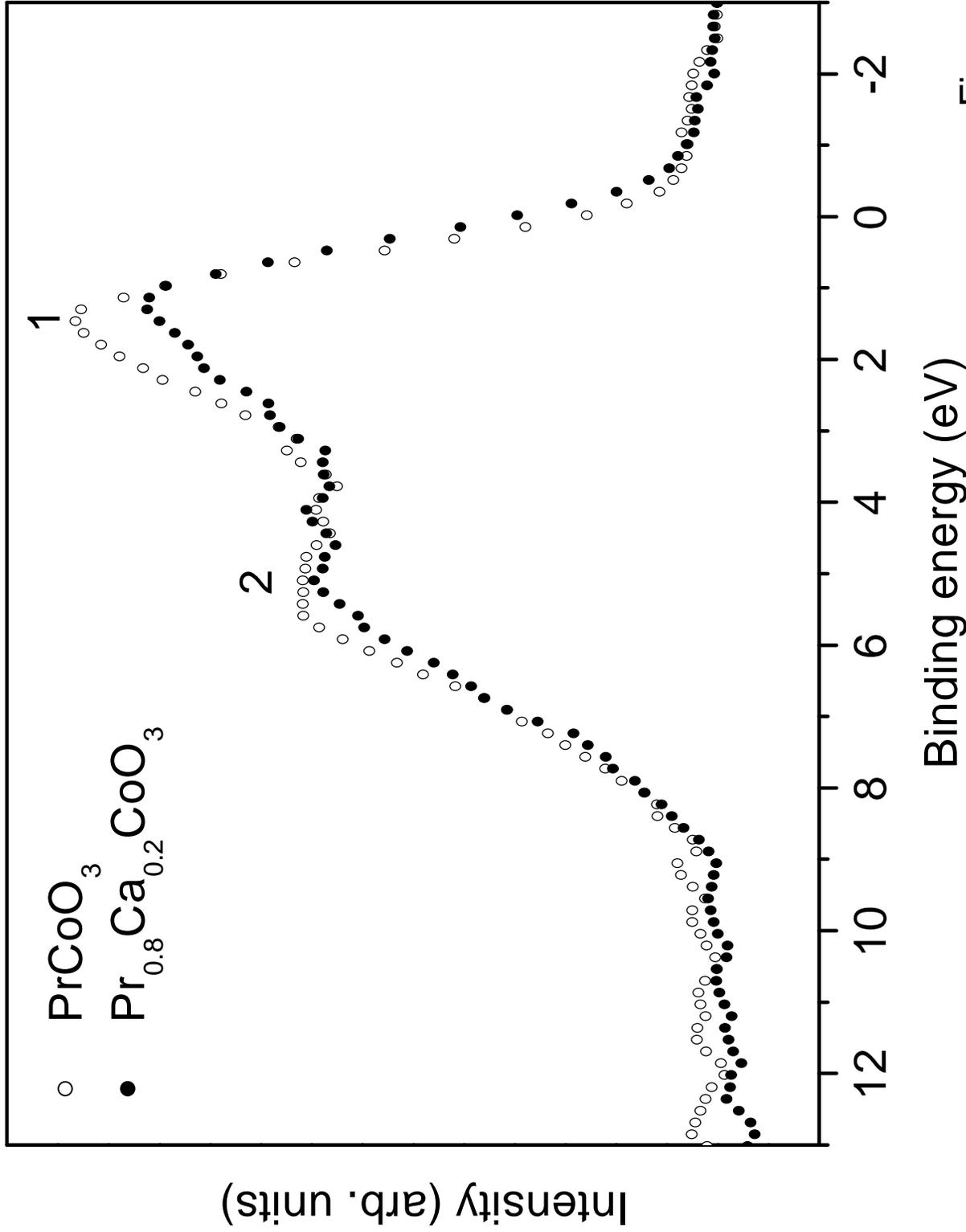

Figure 3
Pandey et al.

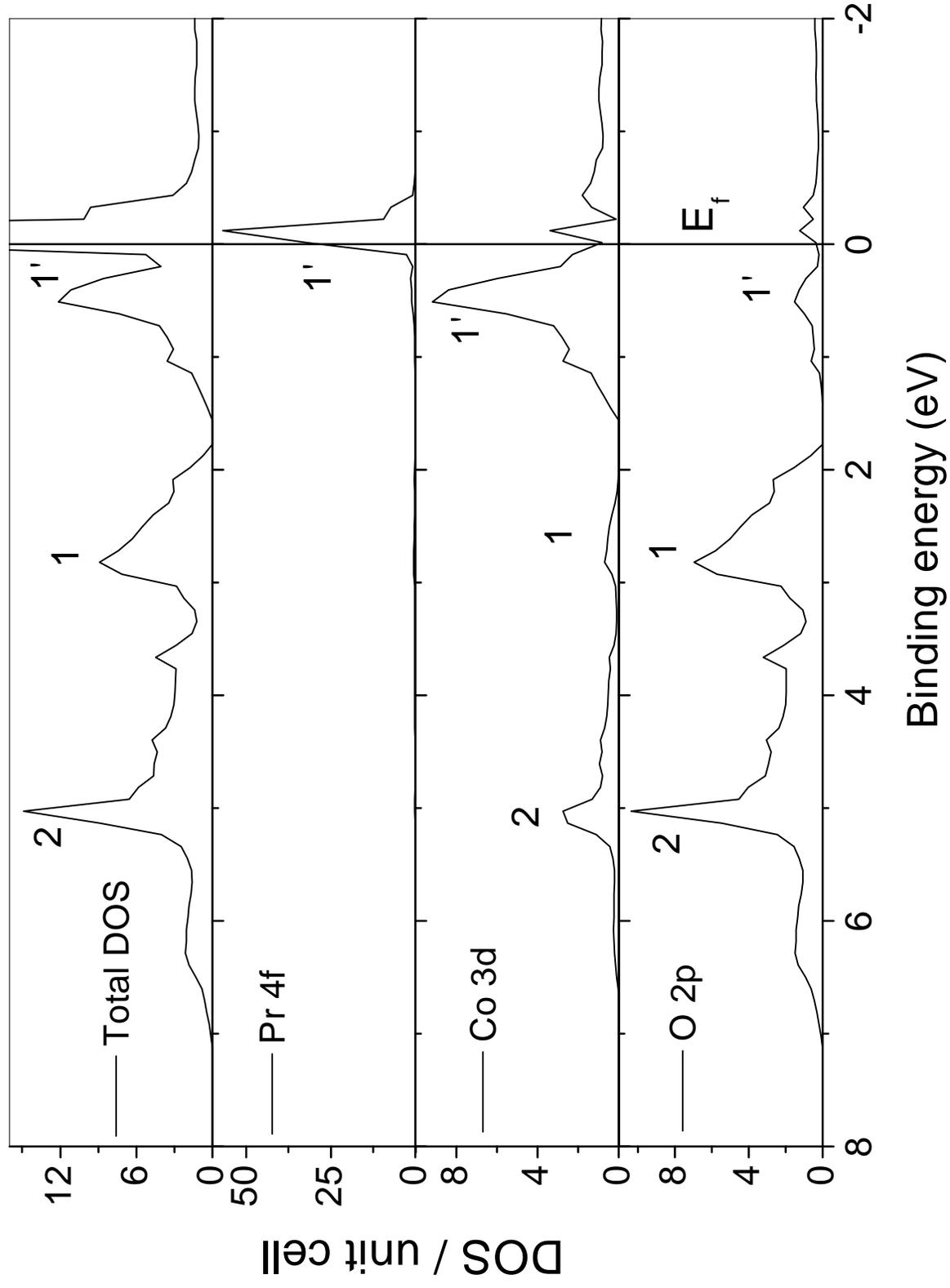

Figure 4
Pandey et al.

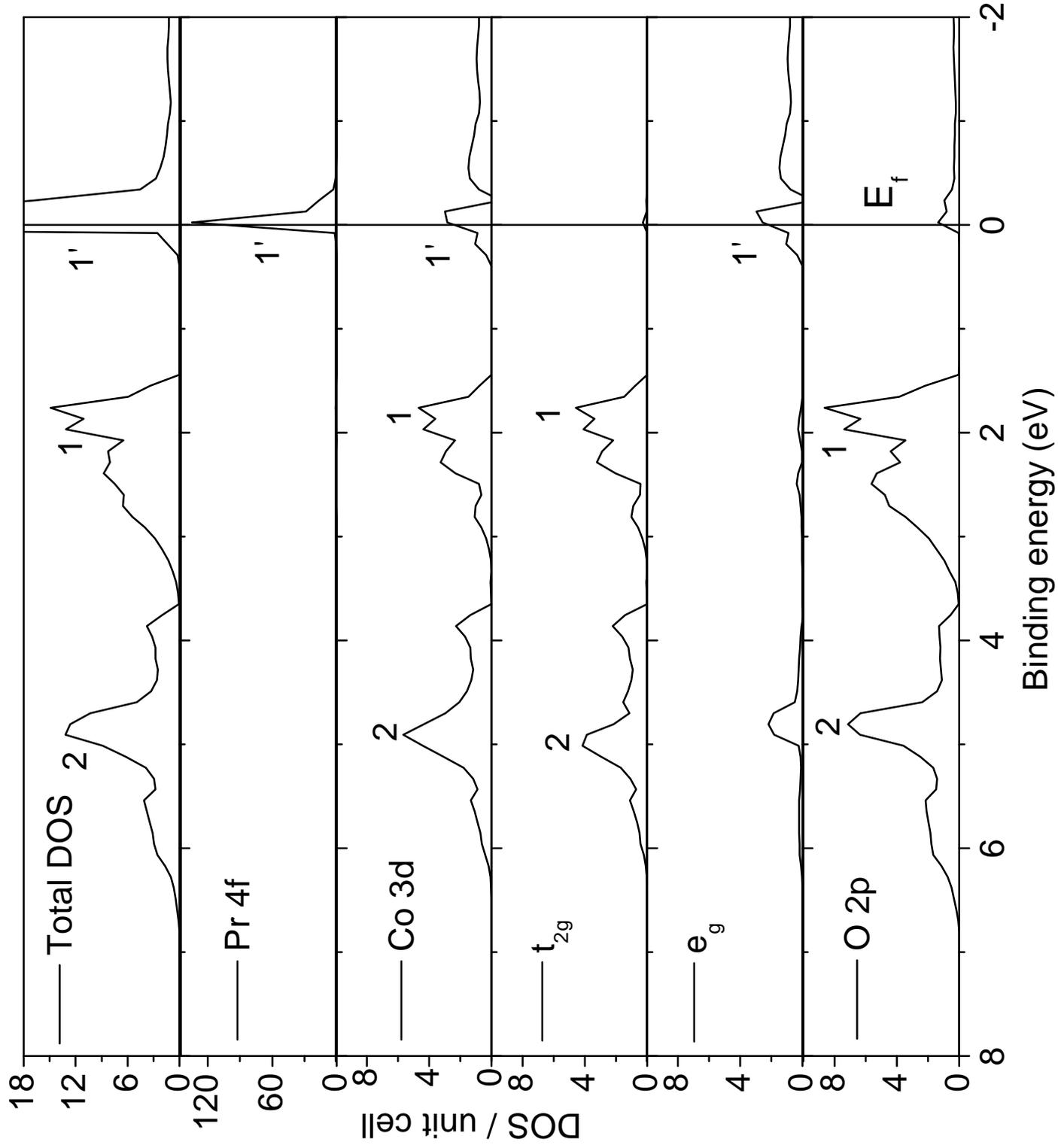

Figure 5
Pandey et al.

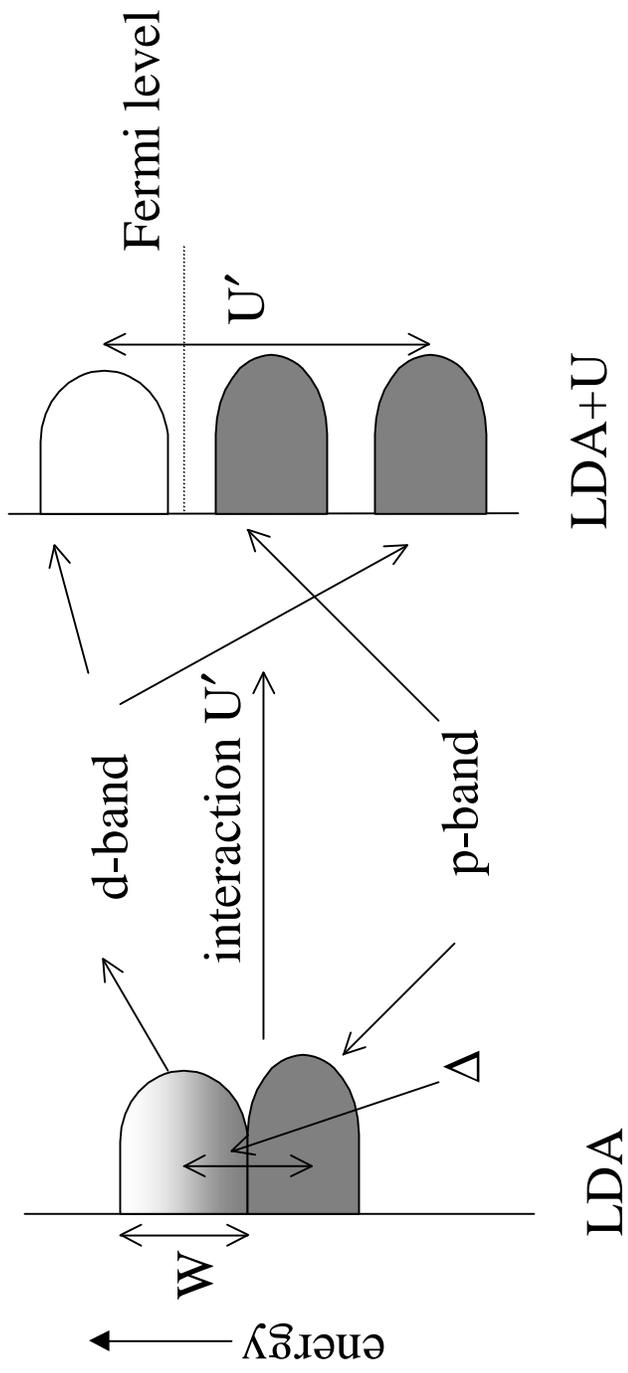

Charge Transfer Insulator

Figure 6
Pandey et al.

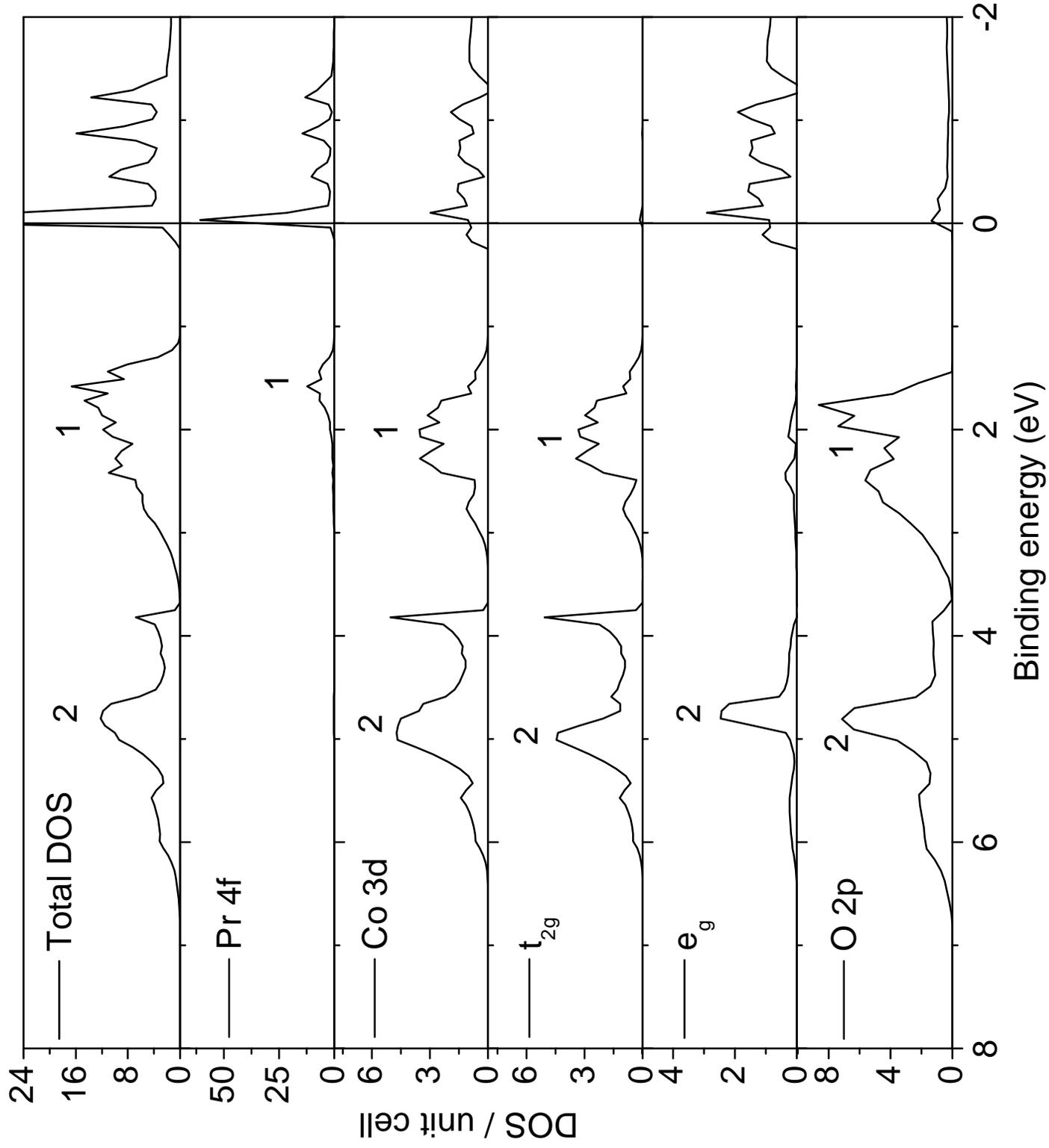

Figure 7
Pandey et al.